\documentclass[onecolumn,showpacs,preprintnumbers,amsmath,amssymb]{revtex4}

\usepackage{graphicx}
\usepackage{dcolumn}
\usepackage{bm}

\begin{document}

\title{Disturbance-free single-pixel imaging camera based on P=0.5 Bernoulli modulation and complementary detection}

\author{Wenlin Gong}
\email{wlgong@suda.edu.cn, gongwl@siom.ac.cn}
\affiliation{School of Optoelectronic Science and Engineering, Soochow University, Suzhou 215006, China}

\date{\today}

\begin{abstract}
We present a technique called single-pixel imaging camera based on complementary detection and optimized Bernoulli modulation (CSPI camera), which can significantly reduce the influence of the disturbance light to single-pixel imaging (SPI). The numerical simulation demonstrates that when the probability of the value ``1" is P=0.5 for each coded pattern, CSPI camera is disturbance-free is still disturbance-free even if the intensity fluctuation of the disturbance light is much larger than the signal's intensity. The reconstruction results of both SPI and differential SPI are also compared. This technique of CSPI camera can dramatically promote real application of single-pixel imaging Lidar.
\end{abstract}


\maketitle

\section{Introduction}

Single-pixel imaging (SPI), as a computational imaging technique, has attracted a lot of attention recently \cite{Cheng,Bennink,Cao,Angelo,Shapiro,Graham-Rowe,Duarte,Edgar,Quero}. At present, there are two typical schematics for SPI \cite{Edgar,Quero}. One is ghost imaging (GI), where the target is illuminated by a series of speckle patterns and the photons reflected from the target are collected onto a single-pixel detector \cite{Cheng,Bennink,Cao,Angelo,Shapiro}. The other is single-pixel camera (SPC), where the target is usually imaged onto a spatial modulation device and the modulated signals are received by a single-pixel detector \cite{Graham-Rowe,Duarte}. In comparison with SPC, more investigations were focused on GI because of its potential applications in remote sensing \cite{Zhao,Erkmen,Wang}, three-dimensional imaging \cite{Sun,Gong,Sun1}, phase imaging \cite{Gong1,Yu,Vinu,Peng}, optical encryption \cite{Clemente,Li}, and microwave imaging \cite{Wang1}. Recently, some works have demonstrated that SPC has more advantages than GI in remote sensing \cite{Edgar,Quero,Mei,Gong2,Sun2}. Firstly, the detection range of GI Lidar is limited by the damage threshold of the modulation device, whereas there is no need for SPC to be considered because the reflection signal is usually weak \cite{Mei}. Secondly, the quality of SPC is better than GI in the same light disturbance environment \cite{Gong2}. Thirdly, the detection process of SPC usual satisfies the standard modeling of compressive sensing, whereas the case of GI is deviated because only the low-frequency information reflected from the target is detected by the single-pixel detector in long-distance imaging \cite{Zhao,Gong,Quero,Sun2}. In addition, the structure's size of SPC is usually smaller than that of GI. Therefore, the schematic of SPC may be more suitable than that of GI in the area of Lidar remote sensing.

Digital micro-mirror device (DMD), as a high-speed light modulator, is widely used for SPC system \cite{Duarte,Edgar,Quero}. By controlling the micro-mirrors of DMD, we can obtain a series of random binary coded patterns with different statistical distribution and the property of coded patterns has a great effect on the quality of SPC \cite{Quero,Wang2,Zhou}. For example, Hadamard coded patterns, as a special case of Bernoulli modulation, can usually obtain better reconstruction result compared with other random Bernoulli coded patterns \cite{Wang2,Zhou}. What's more, because of the modulation property of DMD, the quality of SPC can be also enhanced by complementary detection \cite{Yu1}. However, different from traditional imaging, lots of measurements are required for the image reconstruction of SPI and the intensity fluctuation of the background/disturbance light will lead to the degeneration of SPI in the sampling process especially for the application of SPI Lidar \cite{Gong2,Li1,Deng}. In order to reduce the influence of the background/disturbance light to SPI Lidar, there are two approaches up to now. One is to increase the transmitting energy of Lidar and to decay the intensity of the background/disturbance light by spectral filtering. However, the enhancement degree is usually confined in practical application. The other is based on differential measurement, but it is valid only for the background/disturbance light with gently intensity fluctuation \cite{Yang} and its intensity fluctuation can not be much stronger than the signal's intensity (we will verify it in this manuscript). Furthermore, the intensity fluctuation of the background/disturbance light is usually random in practice. Therefore, it is imperative to develop a method of SPC against the background/disturbance light where its intensity fluctuation is rapidly variable and even is much larger than the signal's intensity. In this paper, we propose a single-pixel imaging camera based on complementary detection and optimized Bernoulli modulation (called CSPI camera) that is disturbance-free when the background/disturbance light illuminating on the DMD plane of CSPI camera is spatially uniform and time-variable. Based on previous SPI research achievement, this work can dramatically quicken real application of SPI Lidar.

\section{Model and image reconstruction}

\begin{figure}[htb]
\centerline{\includegraphics[width=10.0cm]{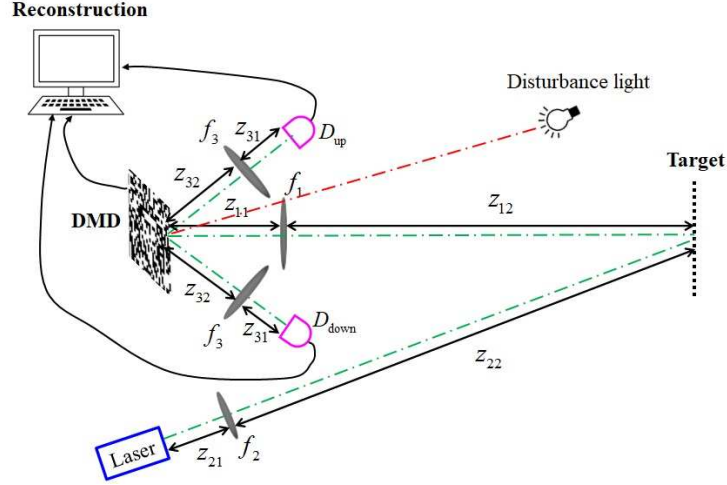}}
\caption{Proof-of-principle schematic of CSPI camera in light disturbance environment. The disturbance light directly enters into the imaging system and uniformly illuminates the DMD, but its intensity is time-variation.}
\end{figure}

Fig. 1 presents the proof-of-principle schematic of CSPI camera in light disturbance environment. The light emitted from a pulsed laser uniformly illuminates the target and the target is imaged onto a DMD by an optical imaging system with the focal length $f_1$. At the same time, the disturbance light directly enters into the imaging system and illuminates the DMD. By controlling the mirrors of the DMD, both the disturbance light and the target's image are modulated, and then the photons reflected by the DMD are collected onto two single-pixel detectors $D_{\rm{up}}$ and $D_{\rm{down}}$ by using another two conventional imaging system with the focal length $f_3$, respectively. According to the property of DMD, the patterns at the plane of the detectors $D_{\rm{up}}$ and $D_{\rm{down}}$ are complementary. We emphasize that the patterns modulated by the DMD obey Bernoulli distribution. In addition, the intensity of the pulsed laser on the target plane is stable, and the disturbance light on the DMD plane is spatially uniform but its intensity is different for each measurement. The intensity $Y_{\rm{up}}^i$ recorded by the detector $D_{\rm{up}}$ can be represented as \cite{Goodman}
\begin{eqnarray}
Y_{\rm{up}}^i  = \int {A^i (x)\left( {I_0 T(x)+I_b^i} \right)} dx+I_{n-up}^i, {\rm{ \ }} \forall _i  = 1 \cdots K,
\end{eqnarray}
where $A^i (x)$ denotes the distribution of the pattern modulated by DMD for the $i$th measurement and $T(x)$ is the intensity reflection function of the target. In addition, $K$ is the total measurement number and $I_0$ is the intensity of the pulsed laser on the target plane. $I_b^i$ and $I_{n-up}^i$ are the intensity of the disturbance light and the detection noise of the detector $D_{\rm{up}}$ for the $i$th measurement, respectively.

Because the detection process is complementary, the intensity $Y_{\rm{down}}^i$ recorded by the detector $D_{\rm{down}}$ can be described as
\begin{eqnarray}
Y_{\rm{down}}^i  = \int {\left( {1-A^i (x)} \right)\left( {I_0 T(x)+I_b^i} \right)} dx+I_{n-down}^i, {\rm{ \ }} \forall _i  = 1 \cdots K,
\end{eqnarray}
here $I_{n-down}^i$ is the detection noise of the detector $D_{\rm{down}}$ for the $i$th measurement.

According to the principle of SPI, the target's image $O_{\rm{SPI}}$ can be reconstructed by computing the correlation function between the pattern's intensity distributions $A^i(x)$ modulated by the DMD and the intensities $Y^i$ recorded by the detector \cite{Cheng,Angelo,Gong2,Quero}
\begin{eqnarray}
O_{\rm{SPI}}(x) = \frac{1}{{K }}\sum\limits_{i =1}^{K}\left( {A^i(x)-\left\langle {A(x)} \right\rangle} \right)Y^i.
\end{eqnarray}
where ${\left\langle {A(x)} \right\rangle} =\frac{1}{{K }}\sum\limits_{s = 1}^{K}A^i(x)$ represents the ensemble average of $A^i(x)$.

For CSPI, the image can be achieved by computing the correlation function between the pattern's intensity distributions $A^i(x)$ and the intensities of $Y_{\rm{up}}^i-Y_{\rm{down}}^i$, namely
\begin{eqnarray}
O_{\rm{CSPI}}(x) = \frac{1}{{K }}\sum\limits_{i =1}^{K}\left( {A^i(x)-\left\langle {A(x)} \right\rangle} \right)Y_{\rm{CSPI}}^i.
\end{eqnarray}
where we denote $Y_{\rm{CSPI}}^i= Y_{\rm{up}}^i-Y_{\rm{down}}^i$, namely
\begin{eqnarray}
Y_{\rm{CSPI}}^i  = I_0 \int {\left( {2A^i (x)-1} \right)T(x)} dx+ I_b^i\int {\left( {2A^i (x)-1} \right)} dx+I_{n-up}^i-I_{n-down}^i, {\rm{ \ }} \forall _i  = 1 \cdots K,
\end{eqnarray}
From Eq. (5), we find that if the speckle patterns $A^i (x)$ obey to Bernoulli distribution (supposed that the probability of the value ``1" is $P$ for each speckle pattern, then the probability of the value ``0" is 1-$P$) and $P$=0.5, then the second item of Eq. (5) is 0 and Eq. (5) can be expressed as
\begin{eqnarray}
Y_{\rm{CSPI}}^i  = I_0 \int {\left( {2A^i (x)-1} \right)T(x)} dx+I_{n-up}^i-I_{n-down}^i, {\rm{ \ }} \forall _i  = 1 \cdots K,
\end{eqnarray}
It means that the disturbance light has no influence to the quality of CSPI in this special case.

In Ref. \cite{Yang}, the optical background noise can be restrained by instant ghost imaging to some extent. To compared with the result of CSPI, we introduce this method into the reconstruction of SPI, which is called differential single-pixel imaging (DSPI) in this manuscript, the reconstruction process can be described as
\begin{eqnarray}
O_{\rm{DSPI}}(x) = \frac{1}{{K-1 }}\sum\limits_{i =1}^{K-1}\left( {A^{i+1}(x)-A^i(x)} \right)Y_{\rm{DSPI}}^i.
\end{eqnarray}
where $Y_{\rm{DSPI}}^i= Y^{i+1}-Y^i$.

In order to evaluate quantitatively the quality of images reconstructed by the methods described above, the reconstruction fidelity is estimated by calculating the peak signal-to-noise ratio (PSNR):
\begin{eqnarray}
{\rm{PSNR} } = 10 \times \log _{10} \left[ {\frac{{(2^p  - 1)^2 }}{{{\rm{MSE} }}}} \right].
\end{eqnarray}
where the bigger the value PSNR is, the better the quality of the recovered image is. For a 0$\sim$255 gray-scale image, $p$=8 and MSE represents the mean square error of the reconstruction images $O_{\rm{rec}}$ with respect to the original object $O$, namely
\begin{eqnarray}
{\rm{MSE} }=\frac{1}{{N_{pix}}}\sum\limits_{i = 1}^{N_{pix}}{\left[ {O_{{\rm{rec }}} (x_i) - O (x_i)} \right]} ^2.
\end{eqnarray}
where $N_{pix}$ is the total pixel number of the image.

\section{Simulated demonstration}

\begin{figure}[htb]
\centerline{\includegraphics[width=8.0cm]{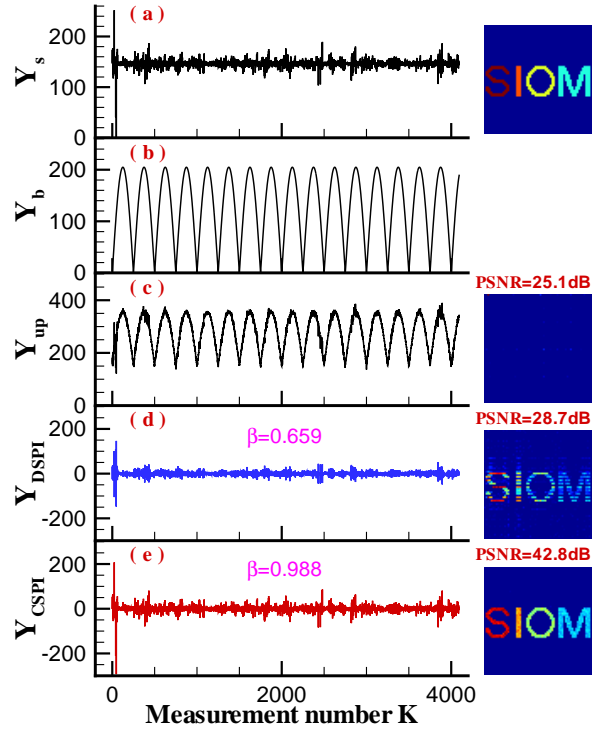}}
\caption{The influence of the disturbance light with sinusoidal modulation to the detection signals and different SPI methods, when the irradiation SNR $\varepsilon$ is 10 dB. (a) The target's detection signal and reconstruction result in the case of no detection noise and no light disturbance; (b) the time-variation intensity of the disturbance light with sinusoidal modulation; (c) the signal recorded by the detector $D_{\rm{up}}$ and corresponding reconstruction result; (d) the detection signal based on the method of DSPI and its reconstruction result; (e) the detection signal based on complementary detection and CSPI reconstruction result.}
\end{figure}

\begin{figure}[htb]
\centerline{\includegraphics[width=10.0cm]{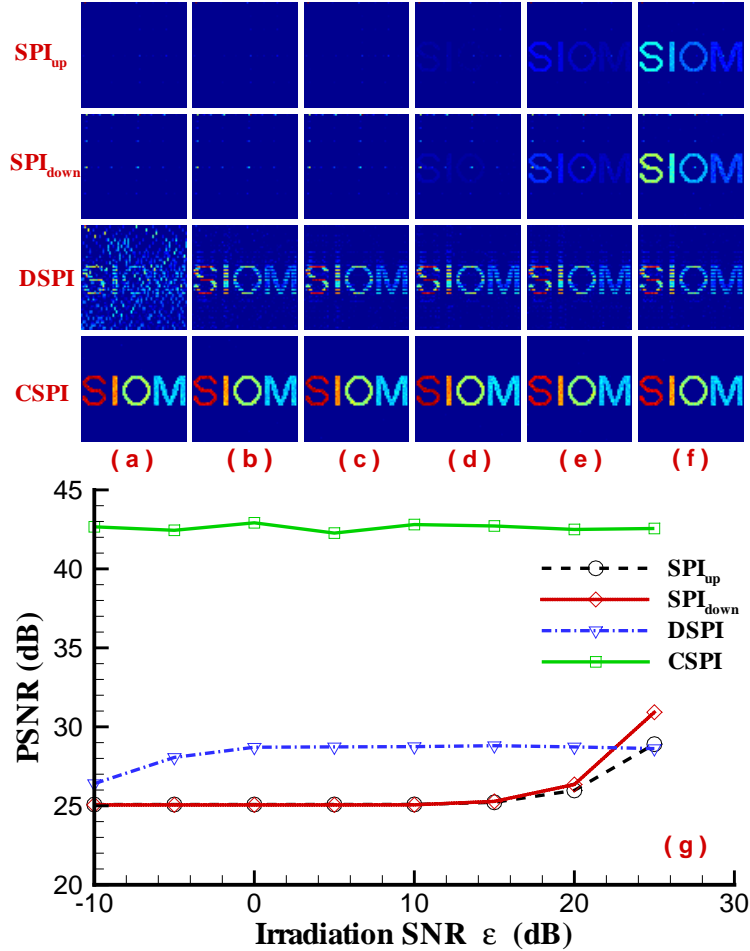}}
\caption{The effect of the intensity of sinusoidal disturbance light on different SPI reconstruction methods. (a) $\varepsilon$=-10 dB; (b) $\varepsilon$=0 dB; (c) $\varepsilon$=10 dB; (d) $\varepsilon$=15 dB; (e) $\varepsilon$=20 dB; (f) $\varepsilon$=25 dB; (g) the curve of PSNR-$\varepsilon$.}
\end{figure}

\begin{figure}[htb]
\centerline{\includegraphics[width=9.5cm]{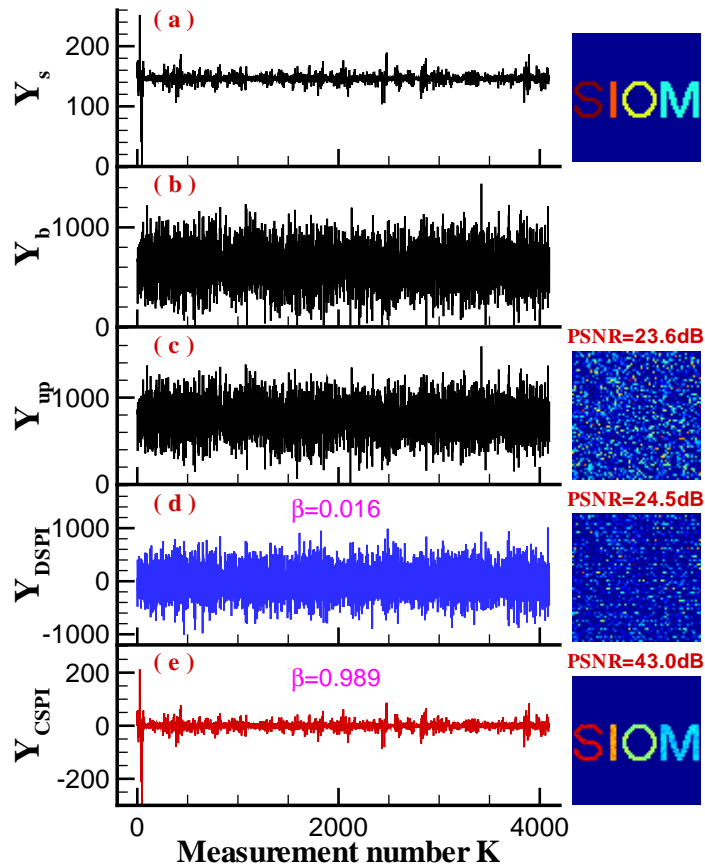}}
\caption{The influence of the disturbance light with Gaussian modulation to the detection signals and different SPI reconstruction methods in the case of irradiation SNR $\varepsilon$=10 dB. The description of (a)-(e) is the same as Fig. 2.}
\end{figure}

\begin{figure}[htb]
\centerline{\includegraphics[width=10.0cm]{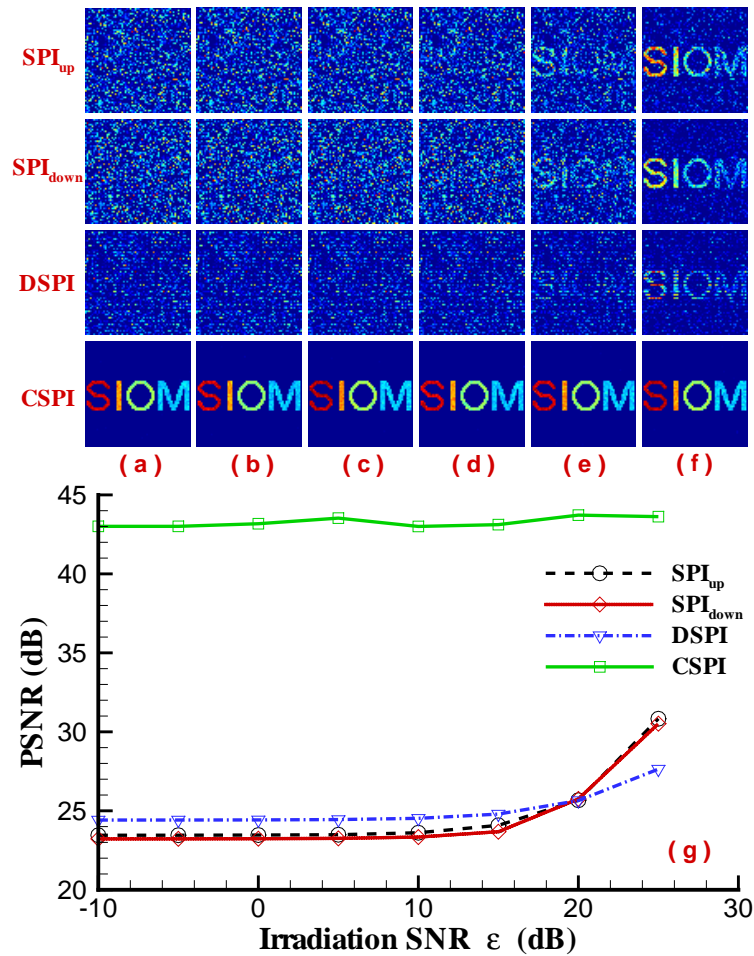}}
\caption{Simulated results of different SPI reconstruction methods in different $\varepsilon$ for the disturbance light with Gaussian statistical distribution. (a) $\varepsilon$=-10 dB; (b) $\varepsilon$=0 dB; (c) $\varepsilon$=10 dB; (d) $\varepsilon$=15 dB; (e) $\varepsilon$=20 dB; (f) $\varepsilon$=25 dB; (g) the curve of PSNR-$\varepsilon$.}
\end{figure}

\begin{figure}[htb]
\centerline{\includegraphics[width=10.0cm]{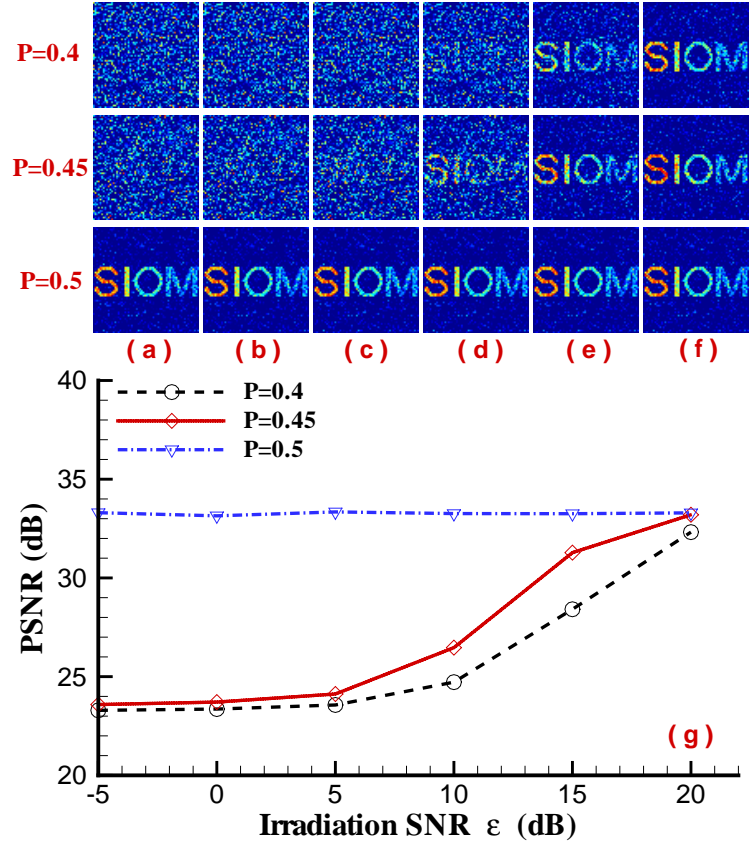}}
\caption{Simulated results of CSPI in different irradiation SNR $\varepsilon$ when the patterns $A^i (x)$ is a random bernoulli distribution and the probability of the value ``1" is P=0.4, 0.45 and 0.5, respectively. (a) $\varepsilon$=-5 dB; (b) $\varepsilon$=0 dB; (c) $\varepsilon$=5 dB; (d) $\varepsilon$=10 dB; (e) $\varepsilon$=15 dB; (f) $\varepsilon$=20 dB; (g) the curve of PSNR-$\varepsilon$.}
\end{figure}

\begin{figure}[htb]
\centerline{\includegraphics[width=12.0cm]{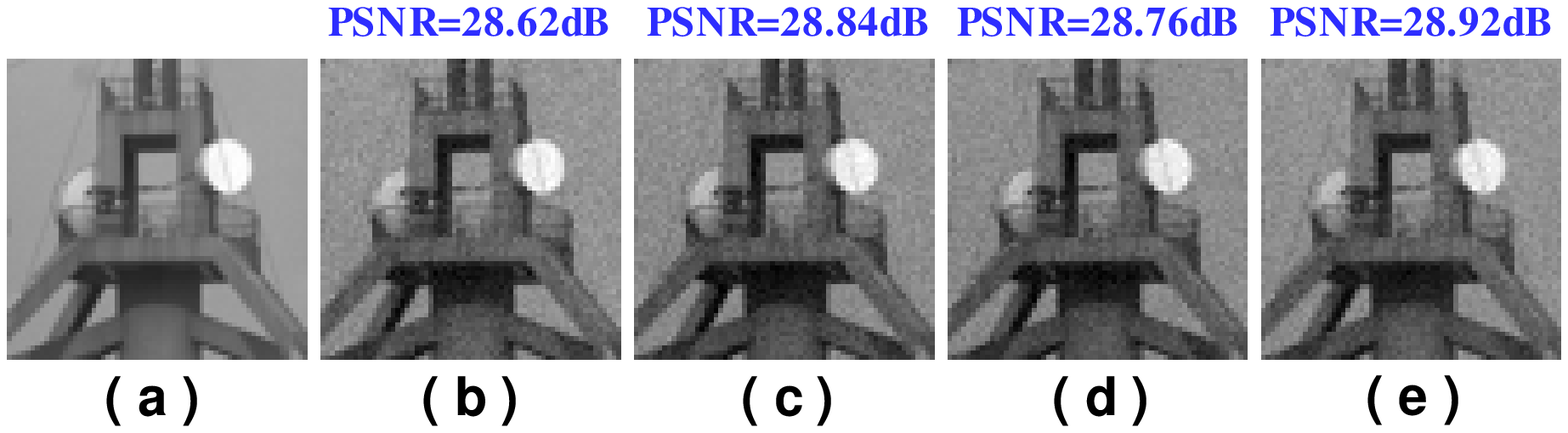}}
\caption{Simulated demonstration of imaging a scene (``tower") in different irradiation SNR $\varepsilon$ when the patterns $A^i (x)$ is the same as Fig. 5 and the detection SNR $\delta$ is 30 dB. (a) the scene; (b) $\varepsilon$=-20 dB; (c) $\varepsilon$=-10 dB; (d) $\varepsilon$=10 dB; (e) $\varepsilon$=20 dB.}
\end{figure}

To verify the concept, the parameters of numerical simulation based on the schematic of Fig. 1 are set as follows: the wavelength of the laser is 532 nm, the transverse size of the patterns at the DMD plane is set as 27.3 $\mu$m and the modulated area of the DMD is 64$\times$64 pixels (one pixel is equal to the pattern's transverse size). In addition, $z_{11}$=$z_{12}$=$z_{21}$=$z_{22}$=200 mm, $f_1$=$f_2$=100 mm. The imaging target, as illustrated in Fig. 2(a), is a gray-scale object (``siom", 64$\times$64 pixels). According to Eq. (1) and Eq. (2), the detection SNR $\delta  = \frac{{\left\langle { \int {A^i (x)I_0 T(x)dx}} \right\rangle }}{{std(I_{n-up}^i )}}$ denotes the signal power to the standard deviation of the noise power ratio, and the irradiation SNR $\varepsilon  = \frac{{I_0}}{{std(I_b^i)}}$ denotes the signal power to the standard deviation of the light disturbance power ratio. The speckle patterns modulated by the DMD obey Bernoulli distribution. In order to improve the sampling efficiency, the Bernoulli speckle patterns are arranged as the type of Hadamard codes (where the position of the value ``-1" is set as 0) and the first pattern (namely the value of all the element is 1) is removed for the demonstration of Fig. 2-Fig. 5. Therefore, the measurement number used for this special case is $K$=4095. What's more, the detection SNR $\delta$ is set as 20 dB. Fig. 2 has given the detection signal and the reconstruction results of SPI/DSPI/CSPI when the disturbance light obeys sinusoidal modulation and the irradiation SNR $\varepsilon$ is 10 dB. When there is no detection noise and no light disturbance in the detection process, both the ideal detection signal and the target's reconstruction image are shown in Fig. 2(a), which is also corresponding to the target's original image. Fig. 2(b) and Fig. 2(c) present the time-variation intensity of the disturbance light and the signal recorded by the detector $D_{\rm{up}}$. It is clearly seen that the target's signal (Fig. 2(a)) is nearly overwhelmed by the disturbance light and traditional SPI reconstruction can not rebuild the target's image (Fig. 2(c)). The detection signals and reconstruction images based on the methods of DSPI and CSPI are displayed in Fig. 2(d) and Fig. 2(e), respectively. By computing the correlation coefficient $\beta$ between the signal $Y_{\rm{s}}$ and the signals $Y_{\rm{DSPI}}$/$Y_{\rm{CSPI}}$\cite{Mei}, the value $\beta$ of DSPI is 0.659 whereas it approaches to 1 for CSPI, which means that the effect of the disturbance light is wiped off by CSPI method. Similar to the result described in Ref. \cite{Yang}, DSPI can improve the reconstruction image to some extent in comparison with traditional SPI for light disturbance with slowly time-variation. However, the target's image can be stably reconstructed by CSPI. Furthermore, the reconstruction results of SPI/DSPI/CSPI in different irradiation SNR $\varepsilon$ are shown in Fig. 3. It is obviously observed that DSPI will be disabled in low $\varepsilon$ whereas the method of CSPI is always valid.

In practice, the intensity fluctuation of the disturbance light is usually random. Similar to Fig. 2, Fig. 4 has presented the detection signals and the reconstruction results of SPI/DSPI/CSPI when the intensity of the disturbance light obeys Gaussian distribution and the irradiation SNR $\varepsilon$ is also 10 dB. Different from the case described in Fig. 2, the signals $Y_{\rm{DSPI}}$ and $Y_{\rm{s}}$ are hardly correlated, thus the image reconstruction is disabled for DSPI. However, CSPI can still stably recover the target's image. Furthermore, Fig. 5 also illustrates simulated results of the influence of the irradiation SNR $\varepsilon$ to SPI/DSPI/CSPI in the condition of the disturbance light with Gaussian statistical distribution. It is clearly seen that SPI is better than DSPI in large $\varepsilon$ and the reconstruction result of CSPI is still perfect even if the irradiation SNR $\varepsilon$ is lower than -10 dB.

The patterns $A^i (x)$ used in Fig. 2-Fig. 5 are a special case of Bernoulli distribution. When the patterns $A^i (x)$ is a random Bernoulli distribution and the probability P of the value ``1" is changed, Fig. 6 has given the simulation results of CSPI in different P and different irradiation SNR $\varepsilon$. Here the simulation parameters are the same as Fig. 5 except for the measurement number $K$=10000. We find that the quality of CSPI is the best in the case of P=0.5 and it will be degraded as the increase of the deviation from the value P=0.5. In addition, compared with the results of CSPI described in Fig. 3 and Fig. 5, the reconstruction results in the case of P=0.5 is worse even if the measurement number $K$ is much larger, which originates from high efficiency in information acquisition for the Bernoulli speckle patterns with Hadamard codes.

In order to validate the applicability of CSPI for complex scenes, Fig. 7 gives a simulation demonstration of imaging a tower captured by our camera (Fig. 7(a)). Using the same simulation parameters as Fig. 5 except for the detection SNR $\delta$=30 dB, the reconstruction results of CSPI are shown in Fig. 7(b)-Fig. 7(e) when the irradiation SNR $\varepsilon$ is -20 dB, -10 dB, 10 dB and 20 dB, respectively, which is similar to the results described in Fig. 5. Therefore, we demonstrate that CSPI is disturbance-free and can be applied to SPI Lidar in the environment of strong background/disturbance light.

\section{Conclusion}

In conclusion, we have proposed a technique called CSPI that can remove the influence of the disturbance light to sing-pixel imaging. We also show that the technique is always valid even if the intensity of the disturbance light is much stronger than the signal's intensity and its intensity fluctuation is random and rapidly variable. This technique can quicken real application of SPI Lidar and is useful to the imaging in the wavebands without cameras.

\section*{Funding}
Natural Science Research of Jiangsu Higher Education Institutions of China (21KJA140001), Aeronautical Science Foundation of China (2020Z073012001), and Startup Funding of Soochow University (NH15900221).


\begin{thebibliography}{99}
\bibitem{Cheng} J. Cheng, and S. Han, ``Incoherent coincidence imaging and its applicability in x-ray diffraction," Phys. Rev. Lett. \textbf{92}, 093903 (2004).
\bibitem{Bennink} R. S. Bennink, S. J. Bentley, R. W. Boyd, and J. C. Howell, ``Quantum and classical coincidence imaging," Phys. Rev. Lett. \textbf{92}, 033601 (2004).
\bibitem{Cao} D. Cao, J. Xiong, and K. Wang, ``Geometrical optics in correlated imaging systems," Phys. Rev. A \textbf{71}, 013801 (2005).
\bibitem{Angelo} M. D. Angelo, and Y. H. Shih, ``Quantum imaging," Laser. Phys. Lett. \textbf{2}, 567-596 (2005).
\bibitem{Shapiro} J. H. Shapiro, and R. W. Boyd, ``The physics of ghost imaging," Quantum Inf. Process. \textbf{11}, 949-993 (2012).
\bibitem{Graham-Rowe} D. Graham-Rowe, ``Digital cameras: Pixel power," Nat. Photonics \textbf{1}, 211-212 (2007).
\bibitem{Duarte} M. F. Duarte, M. A. Davenport, D. Takhar, J. N. Laska, T. Sun, K. F. Kelly, and R. G. Baraniuk, ``Single-pixel imaging via compressive sampling," IEEE Signal Process. Mag. \textbf{25}, 83-91 (2008).
\bibitem{Edgar} M. P. Edgar, G. M. Gibson, and M. J. Padgett, ``Principles and prospects for single-pixel imaging," Nat. Photonics \textbf{13}, 13-20 (2018).
\bibitem{Quero} C. A. Osorio Quero, D. Durini, J. Rangel-Magdaleno, and J. Martinez-Carranza, ``Single-pixel imaging: An overview of different methods to be used for 3d space reconstruction in harsh environments," Rev. Sci. Instruments \textbf{92}, 111501 (2021).
\bibitem{Zhao} C. Zhao, W. Gong, M. Chen, E. Li, H. Wang, W. Xu, and S. Han, ``Ghost imaging lidar via sparsity constraints," Appl. Phys. Lett. \textbf{101}, 141123 (2012).
\bibitem{Erkmen} B. I. Erkmen, ``Computational ghost imaging for remote sensing," J. Opt. Soc. Am. A \textbf{29}, 782-789 (2012).
\bibitem{Wang} C. Wang, X. Mei, P. Long, P. Wang, L. Wang, X. Gao, Z. Bo, M. Chen, W. Gong, and S. Han, ``Airborne near infrared three-dimensional ghost imaging lidar via sparsity constraint," Remote. Sens. \textbf{10}, 732 (2018).
\bibitem{Sun} B. Sun, M. Edgar, R. Bowman, L. Vittert, S. S. Welsh, A. Bowman, and M. J. Padgett, ``3d computational imaging with single-pixel detectors," Science \textbf{340}, 844-847 (2013).
\bibitem{Gong} W. Gong, C. Zhao, H. Yu, M. Chen, W. Xu, and S. Han, ``Three-dimensional ghost imaging lidar via sparsity constraint," Sci. Rep. \textbf{6}, 26133 (2016).
\bibitem{Sun1} M. Sun, M. P. Edgar, G. M. Gibson, B. Sun, N. Radwell, R. Lamb, and M. J. Padgett, ``Single-pixel three-dimensional imaging with time-based depth resolution," Nat. Commun. \textbf{7}, 12010 (2016).
\bibitem{Gong1} W. Gong and S. Han, ``Phase-retrieval ghost imaging of complex-valued objects," Phys. Rev. A \textbf{82}, 023828 (2010).
\bibitem{Yu} H. Yu, R. Lu, S. Han, H. Xie, G. Du, T. Xiao, and D. Zhu, ``Fourier-transform ghost imaging with hard x rays," Phys. Rev. Lett. 1\textbf{17}, 113901 (2016).
\bibitem{Vinu} R. Vinu, Z. Chen, R. K. Singh, and J. Pu, ``Ghost diffraction holographic microscopy," Optica \textbf{7}, 1697-1704 (2020).
\bibitem{Peng} J. Peng, M. Yao, Z. Huang, and J. Zhong, ``Fourier microscopy based on single-pixel imaging for multi-mode dynamic observations of samples," APL Photonics \textbf{6}, 046102 (2021).
\bibitem{Clemente} P. Clemente, V. Duran, V. Torres-Company, E. Tajahuerce, and J. Lancis, ``Optical encryption based on computational ghost imaging," Opt. Lett. \textbf{35}, 2391-2393 (2010).
\bibitem{Li} S. Li, X. Yao, W. Yu, L. Wu, and G. Zhai, ``High-speed secure key distribution over an optical network based on computational correlation imaging," Opt. lett. \textbf{38}, 2144-2146 (2013).
\bibitem{Wang1} X. Wang and Z. Lin, ``Nonrandom microwave ghost imaging," IEEE Transactions on Geoence Remote. Sens. \textbf{56}, 4747-4764 (2018).
\bibitem{Mei} X. Mei, W. Gong, Y. Yan, S. Han, and Q. Cao, ``Experimental research on prebuilt three-dimensional imaging lidar," Chin. J. Lasers \textbf{43}, 0710003 (2016).
\bibitem{Gong2} W. Gong, ``Performance comparison of computational ghost imaging versus single-pixel camera in light disturbance environment," Opt. Laser Technol. \textbf{152}, 108140 (2022).
\bibitem{Sun2} M. J. Sun and J. M. Zhang, ``Single-pixel imaging and its application in three-dimensional reconstruction: A brief review," Sensors \textbf{19}, 732 (2019).
\bibitem{Wang2} C. Wang, W. Gong, X. Shao, and S. Han, ``The influence of the property of random coded patterns on fluctuation-correlation ghost imaging," J. Opt. \textbf{18}, 065703 (2016).
\bibitem{Zhou} C. Zhou, T. Tian, C. Gao, W. Gong, and L. Song, ``Multi-resolution progressive computational ghost imaging," J. Opt. \textbf{21}, 055702 (2019).
\bibitem{Yu1} W. Yu, X. Liu, X. Yao, C. Wang, Y. Zhai, and G. Zhai, ``Complementary compressive imaging for the telescopic system," Sci. reports \textbf{4}, 5834 (2014).
\bibitem{Li1} Z. Li, Q. Zhao, and W. Gong, ``Experimental investigation of ghost imaging in background light environments," J. Opt. \textbf{22}, 025201 (2020).
\bibitem{Deng} C. Deng, L. Pan, C. Wang, X. Gao, W. Gong, and S. Han, ``Performance analysis of ghost imaging lidar in background light environment," Photonics Res. \textbf{5}, 431-435 (2017).
\bibitem{Yang} Z. Yang, W. X. Zhang, M. C. Zhang, R. Dong, and J. L. Li, ``Instant ghost imaging: improving robustness for ghost imaging subject to optical background noise," OSA Continuum \textbf{3}, 391-400 (2020).
\bibitem{Goodman} J. W. Goodman, ``Introduction to fourier optics," (McGraw-Hill, New York, 1968).

\end{thebibliography}
\end{document}